\documentclass[a4paper]{jpconf}
\usepackage{graphicx}
\usepackage{epstopdf}
\usepackage{subfigure}
\usepackage{textcomp}

\def\beq{\begin{equation}}
\def\eeq{\end{equation}}
\def\beqa{\begin{eqnarray}}
\def\eeqa{\end{eqnarray}}
\def\bkR{{\rm I\kern-.17em R}}
\def\bkC{{\rm \kern.24em \vrule width.05em height1.4ex depth-.05ex \kern-.26em C}}

\def\ev{\mbox{eV}}

\def\be{\beta}

\def\frac#1#2{{\textstyle{{#1}\over {#2}}}}
\def\laq{\raise 0.4 ex \hbox{$<$}\kern -0.8 em\lower 0.62 ex\hbox{$\sim$}}
\def\gaq{\raise 0.4 ex \hbox{$>$}\kern -0.7 em\lower 0.62 ex\hbox{$\sim$}}

\def\be{\begin{equation}}
\def\ee{\end{equation}}
\def\ba{\begin{eqnarray}}
\def\ea{\end{eqnarray}}

\def\dalemb#1#2{{\vbox{\hrule height.#2pt
        \hbox{\vrule width.#2pt height#1pt \kern#1pt \vrule width.#2pt}
        \hrule height.#2pt}}}

\def\dalemb#1#2{{\vbox{\hrule height.#2pt
        \hbox{\vrule width.#2pt height#1pt \kern#1pt \vrule width.#2pt}
        \hrule height.#2pt}}}

\def\gtorder{\mathrel{\raise.3ex\hbox{$>$}\mkern-14mu
             \lower0.6ex\hbox{$\sim$}}}
\def\ltorder{\mathrel{\raise.3ex\hbox{$<$}\mkern-14mu
             \lower0.6ex\hbox{$\sim$}}}

\def\lsim{\mathrel{\rlap{\lower4pt\hbox{\hskip1pt$\sim$}}
    \raise1pt\hbox{$<$}}}
\def\gsim{\mathrel{\rlap{\lower4pt\hbox{\hskip1pt$\sim$}}
    \raise1pt\hbox{$>$}}}

\begin{document}


\title{\bf Coupling dark energy with Standard Model states\footnote{Based on a talk delivered by
O.B. at DICE 2008, 22nd - 26th September 2008, Castiglioncello, Italy.}}

\author{M. C. Bento$^{1,2}$, A. E. Bernardini$^{3}$ and O. Bertolami$^{1,4}$}

\address{$^1$ Departamento de F\'\i sica, Instituto Superior T\'ecnico, Avenida Rovisco Pais 1, 1049-001 Lisboa, Portugal. \\

$^2$ Centro de F\'\i sica Te\'orica de Part\'\i culas,  Instituto Superior T\'ecnico, Avenida Rovisco Pais 1, 1049-001 Lisboa, Portugal.  \\

$^3$ Departamento de F\'{\i}sica, Universidade Federal de S\~ao Carlos, PO Box 676, 13595-905, S\~ao Carlos, SP, Brasil.\\

$^4$ Instituto de Plasmas e Fus\~ao Nuclear, Instituto Superior T\'ecnico, Avenida Rovisco Pais 1, 1049-001 Lisboa, Portugal. \\}

\ead {orfeu@cosmos.ist.utl.pt}

\begin{abstract}

{In this contribution one examines the coupling of dark energy to the gauge fields,
to neutrinos, and to the Higgs field.
In the first case, one shows how a putative evolution of the fundamental
couplings of strong and weak interactions via coupling to dark energy
through a generalized Bekenstein-type model may
cause deviations on the statistical nuclear decay Rutherford-Soddy law.
Existing bounds for the weak interaction exclude any significant deviation. For neutrinos,
a perturbative approach is developed which allows for considering viable
varying mass neutrino models coupled to any quintessence-type field. The generalized Chaplygin model is considered as an example.
For the coupling with the Higgs field one obtains an interesting cosmological
solution which includes the unification of dark energy and dark matter. }

\end{abstract}


\section{Introduction}

The nature of the dark energy and its connection with the Standard Model (SM) states is one of the most
intriguing issues related with the negative pressure component required to understand the accelerated expansion of the universe.

A  putative time variation of the electromagnetic
coupling on cosmological time scales from the observation of absorption spectra
of quasars \cite{Webb} was recently reported and  has led to a revival of interest on ideas about the variation of fundamental couplings (see Ref. \cite{Uzan}
for a review). This variation is fairly natural in unification
models with extra dimensions. Moreover, if there is a
variation of the electromagnetic coupling, it is rather natural to
expected variations of other gauge or Yukawa couplings as well.
It is therefore of great interest to confirm these observations and to propose
realistic particle physics models consistent with a cosmological variation of the fundamental couplings.
One  starts analyzing how a putative variation
of the strong and weak couplings due to coupling to dark energy may affect the nuclear decay Rutherford-Soddy law \cite{Bento09}.
In  order to do that one considers the available bounds on the variation of strong and weak couplings and a generalized version of the Bekenstein model \cite{Bekenstein}.

Given that observations suggest that the variation of
the electromagnetic coupling is a late
event in the history of the universe it is  natural to associate it to the observed late
accelerated expansion of the universe \cite{Perlmutter}. This late time acceleration of the
expansion can be driven by a cosmological constant (see \cite{Bento1} and
references therein), by a scalar field with a suitable potential, quintessence \cite{Ratra},
or by a fluid with an exotic equation
of state like the generalized Chaplygin gas (GCG) \cite{Kamenshchik}. For the analysis of the
implications that dark energy may have on the  Rutherford-Soddy law one considers the variation of the strong
and weak couplings in the context of an exponential potential quintessence model and of the GCG model.

The GCG model considers an exotic perfect fluid described by the equation
of state \cite{Kamenshchik}

\beq p = -{A \over \rho^{\alpha}_{ch}}~,
\label{GCGes}
\eeq
\noindent
where
$A$ is a positive constant and $\alpha$ is a constant in the range
$0\leq \alpha \leq 1$. For $\alpha=1$, the
equation of state is reduced to the Chaplygin gas scenario.
The covariant conservation of the energy-momentum tensor for an homogeneous
and isotropic spacetime implies that
\beq
\rho_{ch} = \rho_{ch0}\Bigl[ A_s +
{(1-A_s)  \over a^{3(1+\alpha)}}\Bigr]^{\frac{1}{1+\alpha}}~,
\label{GCGrho}
\eeq
where $A_s=A/\rho_{ch0}^{(1+\alpha)}$,
$\rho_{ch0}$ is the present energy density of the GCG and $a$ the scale factor of the universe. Hence, one can see that at early times
the energy density behaves as matter while at late times
it behaves like a cosmological constant. This dual role
is at the core of the surprising properties of the GCG
model. Moreover, this dependence with the scale factor
indicates that the GCG model can be interpreted as an
entangled admixture of dark matter and dark energy.

This model has been thoroughly studied from
the observational point of view and it is shown to be compatible
with the Cosmic Microwave Background Radiation
(CMBR) peak location and amplitudes \cite{Bento03,Barreiro08}, with SNe
Ia data \cite{Alcaniz,realGCG,Bento05}, gravitational lensing
statistics \cite{Silva},
cosmic topology \cite{Bento06}, gamma-ray bursts \cite{BS06} and variation
of the electromagnetic coupling \cite{Bento07}.
The issue of structure
formation and its difficulties \cite{Sandvik} have been
recently addressed \cite{Bento04b}. Most recent analysis based on CMBR data
indicates that $\alpha < 0.25$ and
$A_s > 0.93 $ \cite{Barreiro08}.


One also studies the coupling of dark energy to neutrinos in the context of the so-called mass
varying neutrino (MaVaN) models \cite{Gu03,Far04,Pas06,Bja08,Ber08b}.
This possibility is particularly interesting since the coupling of neutrinos to the dark energy
scalar field component may lead to a number of significant phenomenological consequences.
If the neutrino mass $m_{\nu}$ is generated by the dynamical value of a cosmologically active
scalar field $\phi$ it would be an evolving quantity.

Hence one considers the possibility that neutrino masses arise from an interaction with the
scalar field that drives the accelerated expansion of the universe.
One considers the MaVaN contribution to the energy conservation equation of the cosmic fluid
as a perturbation that iteratively modifies the background fluid equation of state \cite{Ber08b}.

Dark matter is most often not considered in the formulation of MaVaN models; however,
the possibility of treating dark energy and dark matter in a unified scheme naturally offers this possibility.
The GCG is particularly relevant in this respect, as already mentioned, it is consistent
with known observational constraints.
Our analysis considers the coupling of MaVaN's to the underlying scalar field of the  GCG model.
Since the neutrino contribution is perturbative, one obtains a small deviation from the stability
condition characteristic of the unperturbed GCG equation of state \cite{Ber08b}.


Finally, one also considers a possible coupling between the Higgs boson and dark energy \cite{rosen1}.
In addition to unraveling the mass generation for fermions and electroweak gauge bosons,
the Higgs boson may also be a portal to new physics hitherto hidden in a SM singlet sector \cite{portal}.

The most important consequences of this extension, both cosmological and phenomenological, have been analyzed.
The phenomenological implications can arise in two
ways: mixing of the singlet with the Higgs boson and the possibility of invisible decay of
the Higgs boson into two singlet bosons \cite{portal,mnmsm1,stealth,wells,gabe}. A coupling of the
form $\Phi H^\dagger H$, where $H$ is the Higgs doublet and $\Phi$ is the scalar singlet field, will
generate a mixing between the physical Higgs boson and the singlet
after spontaneous electroweak symmetry breaking. In addition, a quartic coupling like
$\Phi \Phi H^\dagger H$ results in the possibility of Higgs boson decay into a pair of
singlets and, in the case of a non-zero vacuum expectation value of the singlet, a mixing
can also be induced. The invisible decay of the Higgs boson could be
detected through the weak gauge boson fusion at the Large Hadron Collider (LHC) \cite{invisible}

From the cosmological point of view, an ultra-light scalar singlet is stable enough to be
a good dark matter candidate, most often referred to as phion
\cite{stealth,gabe,zee1,mcdonald,zee2,burgess}. Bounds on the phion-Higgs coupling can be obtained from the constraints on the phion relic abundance.

One repeats here the analysis of Ref. \cite{rosen1} on the consequences of identifying the scalar singlet
not only with the dark matter, but also with the dark energy field, responsible for the recent stage
of accelerated expansion of the universe.
It is shown that it is possible to obtain, under conditions,
an unified picture of dark matter and dark energy,
where dark energy is the zero-mode classical
field rolling the usual quintessence potential and the dark matter candidate
is the quantum excitation (particle) of the field, which is generated in the universe due to its coupling to the Higgs boson.


\section{Coupling to Gauge Fields}


As an example of the coupling of dark energy to gauge fields, one considers
the impact that this putative coupling might
have on the well established experimental fact that the nuclear decay rate
of any nuclei is described by the statistical Rutherford-Soddy law:
\beq
{dN\over dt} =-\lambda N~~,
\eeq
where $\lambda={1\over \tau_0}$, $\tau_0$ being the nuclei's lifetime.
For a constant $\lambda$, which reflects the fact that the rate of transformation
of an element is constant under all conditions
\cite{RCE30}, one obtains the
well known exponential decay law:
\beq
N(t)=N_i e^{-\lambda (t-t_i)}~~,
\eeq
where $N_i$ is the nuclei's number counting at $t_i$.
This statistical law is a consequence of radiative processes of electromagnetic,
strong and weak nature that take place within the nuclei. Thus,
$\tau_0$ is related to the amplitude of the relevant decay process being
therefore a function of order $\gamma$ of the coupling constant $\alpha_g$:
\beq
\tau_0=A \alpha_g^\gamma~~,
\eeq
with $\alpha_S(E) = g_s^2(E)/\hbar c$ for the strong interactions, $g_s$ being the strong coupling constant, $\alpha_W=G_F  m_p^2 c/\hbar^3$ the weak coupling constant, $G_F$
being Fermi's constant and $m_p$ the proton mass. Parameter $A$ is a constant related to a specific
process, suitable integration over the phase space, binding energy, quark masses, etc.

On the other hand, one expects that if dark energy couples with the whole
gauge sector of the Standard Model, this coupling can be modelled by
a generalization of the
so-called Bekenstein model \cite{Bekenstein}
\beq
{\cal L}_{gauge} =-{1  \over 16\pi} f\left(\phi\right)F_{\mu\nu}^a F^{\mu\nu a}~,
\eeq
where $f(\phi)$ is an arbitrary function of the dark energy field, $\phi$, and $F_{\mu\nu}^a$ the gauge field strength.
Given that the variation of the
gauge couplings is presumably small (c.f. below), one expands this function to first order
\beq
f\left(\phi\right)={1  \over \alpha_{i}}\left[1+\zeta \left({\phi-\phi_i \over M}\right)\right]~,
\eeq
where $\alpha_{i}$ is the initial value of the gauge structure constant, $\zeta$ is a constant,
$\phi_i$ is the initial value of the quintessence field
and $M$ a characteristic mass scale of the dark energy model.
We point out that a model with a quadratic variation is discussed in Ref.
\cite{OP08}. For the linear model it follows that
\beq
\left[f\left(\phi\right)\right]^{-1} \simeq\alpha_{i}\left[1-\zeta
\left({\phi-\phi_i \over M}\right)\right]~.
\label{eq: functionf}
\eeq

Thus, the gauge coupling evolution is given by
\beq \
\alpha_g =\left[f\left(\phi\right)\right]^{-1} \simeq \alpha_{i}\left[1-\zeta
\left({\phi-\phi_i \over M}\right)\right]~,
\label{alpha2}
\eeq
\noindent
hence, for its variation, one obtains
\beq {\Delta\alpha_g  \over \alpha_g}=\zeta\left({\phi-\phi_i \over M}\right)~.
\label{delta2}
\eeq

For the electromagnetic interaction one should take into account the
Equivalence Principle limits, which implies \cite{Equivalence}
\beq
\zeta \leq 7 \times 10^{-4}~.
\label{zetaF}
\eeq
\noindent
Of course, this bound does not apply to
short range interactions. However, as will be seen, this parameter must be
constrained in order to ensure that strong and weak gauge coupling
do not change significantly
so to ensure the experimental standing of the Rutherford-Soddy law.

In what follows one considers only strong and weak decays given that the effects of a
putative variation of the electromagnetic coupling have been the subject
of various studies (see e.g. Ref. \cite{Bento04a} and references therein).
Thus, for strong and weak decays, keeping quark masses unchanged and
disregarding the running of the couplings, one can write for $\tau$:
\beq
\tau_{DE}=A f(\phi)^{-\gamma}~~,
\eeq
and since $\zeta << 1$
\beq
\tau_{DE}\simeq \tau_0\left[1-\gamma \zeta_{S,W} \left({\phi_0-\phi_i \over M}\right)\right]~~,
\eeq
depending on whether the process is driven by strong or weak interactions,
where $\phi_0$ refers to the value of the dark energy scalar field at present.
Introducing this expression into the rate of change of atoms
\beq
{d N \over dt}\simeq -{N\over \tau_0}\left[1+\gamma \zeta_{S,W} \left({\phi_0-\phi_i \over M}\right)\right]~~,
\eeq
from which follows that \cite{Bento09}
\beq
N(t)\simeq N_i \exp\left[-\lambda(\Delta t +\gamma \zeta_{S,W} I)\right]~~,
\eeq
where
\beq
I=\int_{t_i}^{t_0} \left({\phi_0-\phi_i \over M}\right) dt~~,
\label{integral}
\eeq
and $\Delta t=t_0-t_i$.

One considers now the most stringent bounds on the variation of the gauge couplings
(latest bounds on the variation of the electromagnetic coupling
can be found in \cite{Bento04a} and references therein).
In what concerns the strong interaction, considerations on the
stability of two-nucleon systems yield the bound \cite{Uzan}
\beq
\left|{\Delta \alpha_S \over \alpha_S}\right|< 4\times 10^{-2}~~.
\label{boundstrong}
\eeq

On its hand, for weak interactions, considerations on the $\beta$-decay rates of $^{187}Re$ lead to the bound \cite{re,dent}
\beq
{\Delta \alpha_W \over \alpha_W} < 3 \times 10^{-7}~~.
\label{boundweak}
\eeq


In order to make quantitative predictions concerning the level of validity of the
statistical Rutherford-Soddy decay law, one examines two observationally
viable dark energy models.


For simplicity, one starts considering the exponential potential
$V(\phi)=V_o e^{-\beta\phi /M_P }$, in the case that the contribution of the scalar field
is dominant, there is a family of solutions \cite{Ferreira}
\beqa
\phi(t) &= &\phi_0+ {2 \over \beta} \ln (t M_P)~,\\
\qquad \phi_0 &=& {2  \over \beta} \ln\left({V_o\beta^2  \over 2M_P^4(6-\beta^2)} \right)~,\\
\qquad \rho_\phi &\propto & {1 \over a^{\beta^2}}
\qquad a \propto t^{2/\beta^2} ~~.
\label{eq: jhsoln}
\eeqa
in terms of the scale factor of the universe. For $\beta < \sqrt{6}$, $\phi_0$ can always be chosen
to be zero by redefining the origin of $\phi$, in which case
$V_0=\frac{2}{\beta^2}(\frac{6}{\beta^2}-1)M_P^4$. The parameters of
this model can be set to satisfy all
phenomenological constraints (see e.g. last reference in \cite{Ratra}).


One also considers the GCG model. Following Refs. \cite{realGCG,Ber08b}, one describes the GCG through a real scalar
field. One starts with the Lagrangian density
\beq
{\cal L}={1 \over 2}\dot\phi^2-V(\phi)~,
\label{realL}
\eeq
\noindent
where the potential for a flat, homogeneous and isotropic universe has the following form
\beq
V={1\over 2} A_s^{1\over 1+\alpha}\rho_0\Bigl[\cosh(\beta \phi)^{\frac{2}{\alpha+1}}
+\cosh({\beta\phi})^{-{2\alpha \over \alpha+1}}\Bigr]~,
\label{realV}
\eeq
\noindent
where $\beta=3(\alpha+1)/2$.

For the energy density of the field one has
\beq
\rho_\phi=\frac{1}{2}\dot\phi^2+V(\phi)=\Bigl[ A +
{B \over a^{3(1+\alpha)}}\Bigr]^{\frac{1}{1+\alpha}}=\rho_{ch}~.
\label{realRHO}
\eeq

The Friedmann equation can be written as
\beq
H^2={ 8 \pi G \over 3} \rho_{\phi}
\eeq
and thus
\beq
H(t)=H_0 \Omega_\phi^{1/2}
\label{H}
\eeq
where $\Omega_\phi$ can be written as
\beq
\Omega_\phi=\left[A_s+{(1-A_s)\over a^{3(1+\alpha)}}\right]^{1\over 3(1+\alpha)}~.
\eeq

Using Eq.~(\ref{GCGes}) to obtain the pressure and integrating one obtains
\beq
\phi (a)=-{1\over 2\beta}\ln\left[{\sqrt{1-A_s(1-a^{2\beta})}-\sqrt{1-A_s}\over
\sqrt{1-A_s(1-a^{2\beta})} + \sqrt{1-A_s}}\right]~.
\label{integratedPHI}
\eeq


The integral $I$, Eq. (\ref{integral}), has been computed for the quintessence
model with an exponential potential (Model I) and
for the GCG model (Model II) and the implications for the
variation of the Rutherford-Soddy decay law for
strong and weak interactions have been examined. For Model I one has set $\beta^2 = 2$ and that
at present the scalar field contribution to the energy density
is $70\%$ of the critical density. For
Model II, one has considered the values $\alpha =0.2$, $A_s= 0.95$,
and used the relationship $1+z=1/a$ to express the integral $I$ in terms of the red-shift.

For the strong interaction decay one considers the $^{238}_{92}U$
nucleus for which $\tau_0= 4.468 \times 10^9$ years. For the
weak interaction decay, oone uses data from the $\beta^-$ decay of the $^{187}Re$ for
which $\tau_0= 4.558 \times 10^9$ years. Using bounds (\ref{boundstrong}), (\ref{boundweak}),
and the integration interval equal to $\tau_0$
we compute $\zeta_{S,W}$ and the respective relative deviation from the Rutherford-Soddy law  $N/N_{RS}$
for the case of weak and strong
interactions. Our results are shown in Table 1 for $\gamma=2$.
\begin{table*}[ht]
\centering
\begin{tabular}{c c c c c c c c}
\hline
\hline
& I (yrs) & $\mid \zeta_S\mid$ &$\mid\zeta_W\mid$ &
 $\mid{N\over N_{RS}}\mid_S -1\mid$ & $\mid{N\over N_{RS}}\mid_W -1\mid$\\
\hline
\\
Model I  & $1.22\times 10^{9}$ & $  7.1\times 10^{-2}$& $5.24 \times 10^{-7}$&
$3.8\times 10^{-2}$ & $2.79 \times 10^{-7}$  \\
\\
Model II & $1.08\times 10^{9}$ & $2.27\times 10^{-1} $& $1.7 \times 10^{-6} $&
$1.04 \times 10^{-1}$&  $8.1 \times 10^{-7}$
   \\
\\
\hline
\end{tabular}
\end{table*}

\section{Coupling to Neutrinos}

The existence of a cosmological neutrino background is a firm prediction of the
cosmological standard model, hence any hint about this component of the universe energy density is quite relevant.

The neutrino energy density and pressure are expressed through a Fermi-Dirac distribution function without a
chemical potential, $f(q)$, where $q \equiv \frac{|\mbox{\boldmath$p$}|}{T_{\nu 0}}$, and sub-index $0$ denotes present values.
Assuming a flat Friedman-Robertson-Walker cosmology with $a_{0} = 1$, one has
\begin{eqnarray}
\rho_{\nu}(a, \phi) &=&{T^{4}_{\nu 0} \over \pi^{2}\,a^{4}}
\int_{_{0}}^{^{\infty}}{\hspace{-0.3cm}dq\,q^{2}\,\left(q^{2} + {{m^{2}(\phi)\,a^{2}}\over{T^{2}_{\nu0}}}\right)^{1/2}\hspace{-0.1cm}f(q)},\\
p_{\nu}(a, \phi) &=&{T^{4}_{\nu 0} \over 3\pi^{2}\,a^{4}}\int_{_{0}}^{^{\infty}}{\hspace{-0.3cm}dq\,q^{4}\,
\left(q^{2}+{{m^{2}(\phi)\,a^{2}}\over{T^{2}_{\nu 0}}}\right)^{1/2}\hspace{-0.1cm} f(q)}.~~~~ \nonumber
\label{gcg01}
\end{eqnarray}
By observing that
\begin{equation}
m_{\nu}(\phi) {\partial \rho_{\nu}(a, \phi) \over \partial m_{\nu}(\phi)} = \left(\rho_{\nu}(a, \phi) - 3 p_{\nu}(a, \phi)\right),
\label{gcg02}
\end{equation}
and from Eq.~(\ref{gcg01}), one obtains the energy-momentum conservation for the neutrino fluid
\begin{equation}
\dot{\rho}_{\nu}(a, \phi) + 3 H (\rho_{\nu}(a, \phi) + p_{\nu}(a, \phi)) =
\dot{\phi} {d m_{\nu}(\phi) \over d \phi} {\partial \rho_{\nu}(a, \phi) \over \partial m_{\nu}(\phi)},
\label{gcg03}
\end{equation}
where $H = \dot{a}/{a}$ is the expansion rate of the universe and the {\em dot} denotes differentiation with respect to cosmic time.

The coupling between cosmological neutrinos and the scalar field as specified in Eq.~(\ref{gcg02})
is restricted to times when neutrinos are non-relativistic (NR), i. e.
$\frac{\partial \rho_{\nu}(a, \phi)}{\partial m_{\nu}(\phi)} \simeq n_{\nu}(a) \propto{a^{-3}}$ \cite{Far04,Bja08,Pec05}.
On the other hand, as long as neutrinos are relativistic ($T_{\nu}(a) = T_{\nu 0}/a >> m_{\nu}(\phi(a))$),
the decoupled fluids evolve adiabatically since the strength of the coupling is suppressed by
the relativistic increase of pressure ($\rho_{\nu}\sim 3 p_{\nu}$).

Treating the system of NR neutrinos and the scalar field as a single unified fluid (UF) which adiabatically
expands with energy density $\rho_{UF} = \rho_{\nu} + \rho_{\phi}$ and pressure $p_{UF} = p_{\nu} + p_{\phi}$ leads to
\begin{equation}
\dot{\rho}_{UF} + 3 H (\rho_{UF} + p_{UF}) = 0~~ \Rightarrow ~~\dot{\rho}_{\phi} + 3 H (\rho_{\phi} + p_{\phi}) = -\dot{\phi}{d m_{\nu} \over d \phi} {\partial \rho_{\nu} \over \partial m_{\nu}},
\label{gcg06}
\end{equation}
where the last step is derived from the substitution of Eq.~(\ref{gcg03}) into Eq. (\ref{gcg06}).

It is well known that the relative contribution of the energy densities components of the universe
with respect to the one of the dark energy is on its own a problem.
The assumptions proposed in Ref. \cite{Far04} and subsequently developed elsewhere \cite{Bja08,Pec05,Bro06A,Tak06}
introduce a stationary condition (SC) which allows circumventing the coincidence problem for cosmological neutrinos,
by imposing that the dark energy is always diluted at the same rate as the neutrino fluid, that is,
\begin{equation}
{d V(\phi) \over d {\phi}} = - {d m_{\nu} \over d \phi} {\partial \rho_{\nu} \over \partial m_{\nu}}.
\label{gcg07}
\end{equation}
This condition introduces a constraint on the neutrino mass since it promotes it into a dynamical quantity,
as indicated in Eq.~(\ref{gcg06}).
In this context, the main feature of the scenario of Ref. \cite{Far04} is that, in what concerns to dark sector,
it is equivalent to a cosmological constant-like equation of state and an energy density that is as a function
of the neutrino mass \cite{Ber08b}.

At our approach, the effect of the coupling of the neutrino fluid to the scalar field fluid is quantified by a
linear perturbation $\epsilon \phi$ ($|\epsilon| << 1$) such that $\phi \rightarrow \varphi \approx (1 + \epsilon) \phi$.
It then follows a novel equation for energy conservation
\begin{equation}
\ddot{\varphi} + 3 H \dot{\varphi} + {d V({\varphi}) \over d {\varphi}} = -{d m_{\nu} \over d \varphi} {\partial \rho_{\nu} \over
\partial m_{\nu}}.
\label{gcg10}
\end{equation}
After some straightforward manipulation one obtains for the value of the coefficient of the perturbation \cite{Ber08b,portal}
\begin{eqnarray}
\epsilon  \simeq {-{d m_{\nu} \over d \phi} {\partial \rho_{\nu} \over \partial m_{\nu}} \over \left[\phi^{2} {d \over d {\phi}}\left({1\over\phi} {d V(\phi) \over d {\phi}}\right)\right]},
\label{gcg13}
\end{eqnarray}
for which the condition $|\epsilon| << 1$ is required.
Upon fulfilling all known phenomenological requirements, the above result allows for addressing a wide
class of scalar field potentials and related equations of state for various candidates for the dark sector
(dark energy and dark matter), which through the SC would be incompatible with realistic neutrino mass generation models.

In order to verify under which conditions Eq.~(\ref{gcg07}) agrees with the proposed perturbative approach for a
given background equation of state, the coefficient of the linear perturbation should be given by
\begin{eqnarray}
\epsilon  \simeq {{d V(\phi)\over d \phi}\over \left[\phi^{2}
{d \over d {\phi}}\left({1\over \phi}
{d V(\phi)\over d {\phi}}\right)\right]}~~~~ (|\epsilon| << 1).
\label{gcg16}
\end{eqnarray}
This means that one must search for a neutrino mass dependence on the scale factor for which the above condition is satisfied.
Thus, once one sets the equation of state for the dark sector, there will be a period at late times for
which the SC and the perturbative approach match.
In particular, this feature can be reproduced by the GCG equation of state.

\begin{figure}
\vspace{-2.3 cm}
\includegraphics[width=14cm]{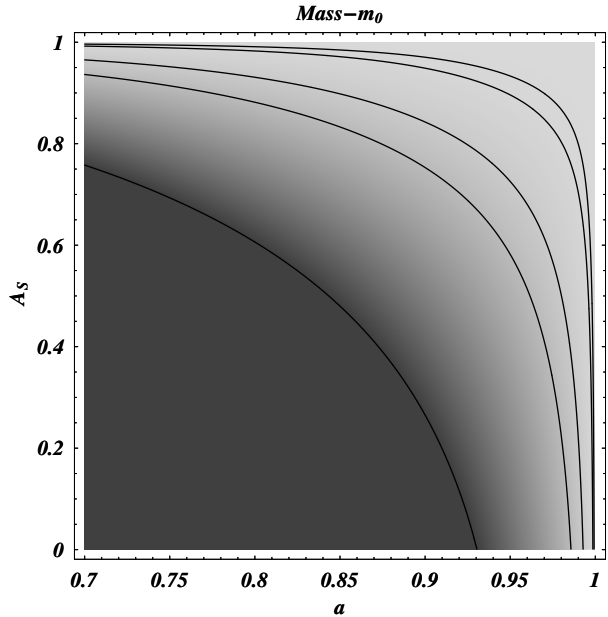}
\vspace{-2.3 cm}
\caption{Present-day values of the neutrino mass $m_{0}$ and the corresponding values of $a_{NR}$
for which the transition between the NR and UR regimes takes place in the GCG
scenario with $\alpha = 1/2$ and variable $A_{s}$.
The increasing {\em graylevel} corresponds to increasing values of $m_{0}$, for the boundary values
$m_{0} = 0.05\, eV,\,0.1\, eV,\,0.5\, eV,\,1\, eV,\, 5\, eV$, respectively.}
\label{fGCG-03}
\end{figure}

Given a potential, the explicit dependence of $m_{\phi}$ on the scale factor can be immediately obtained from Eq.~(\ref{gcg07}).
Furthermore, it is necessary to determine for which values of the scale factor the neutrino-scalar field
coupling becomes important.
For convenience one sets the value of $a = a_{NR}$ for which $\rho_{\nu,NR} = \rho_{\nu,UR}$ holds, usually
established by the condition of $m_{\nu}\gsim T_{\nu}$, that parameterizes the transition between the NR
and the ultra-relativistic (UR) regime.
In fact, this takes place when
\begin{equation}
m_{\nu}(a) = m_{0}(\phi_{0}/\phi(a))^{n} = \chi {T_{\nu, 0} \over a}
\label{gcg33}
\end{equation}
where $\chi$ is a numerical factor estimated to be about $\chi \simeq 94$ considering that
$\rho_{\nu}/\rho_{\mbox{\tiny Crit}} = m_{0}\,[eV]/(94\,h^{2}\,[eV])$, where $h$ is the value
of the Hubble constant in terms of $100\, km\, s^{-1}\,Mpc^{-1}$.
Such a correspondence between $a_{NR}$ and $m_{0}$ is illustrated in the Fig.~\ref{fGCG-03} for the case of $\alpha = 1/2$.
Considering the whole set of parameters that characterize the background fluid, one notices that
it is rather difficult to see that the maximal value assumed by the $\epsilon$ parameter corresponds to its present-day value.

One observes that the interval of parameters $A_{s}$, $m_{0}$ and eventually $\alpha$,
for which our approximation can be applied ($\epsilon < 1$), is valid for $a > a_{NR}$ and severally
constrained by the condition $a_{NR}\lsim 1$.
For  values of $A_{s}$ ($ 0.7 \lsim A_{s} \lsim 1)$  \cite{Bento03,Bento05} one finds that $0.1 < \epsilon \lsim 1$.
Just under quite special circumstances the usual SC and the perturbative contribution of MaVaN's match.
In the original MaVaN scenario \cite{Far04}, the SC corresponds to the adiabatic solution
($H^{2} \ll \mbox{d}^{2}V/\mbox{d}\phi^{2}$) of the scalar field equation of motion.
In this case, the kinetic energy terms of the scalar field can be safely neglected.
The consistency of our perturbative scenario with the stationary condition can be achieved
only when the kinetic energy contribution is not relevant at late times.

For $A_{s} =0$, the GCG behaves always as matter, whereas for $A_{s} =1$, it behaves always as a cosmological constant.
Consequently, it is natural that the relevance of the kinetic energy term at present times is suppressed
when the parameter $A_{s}$ gets close to unity, which further ensures the agreement between the
perturbative approach and the SC analysis.

\begin{figure}
\vspace{-0.5 cm}
\includegraphics[width=14cm]{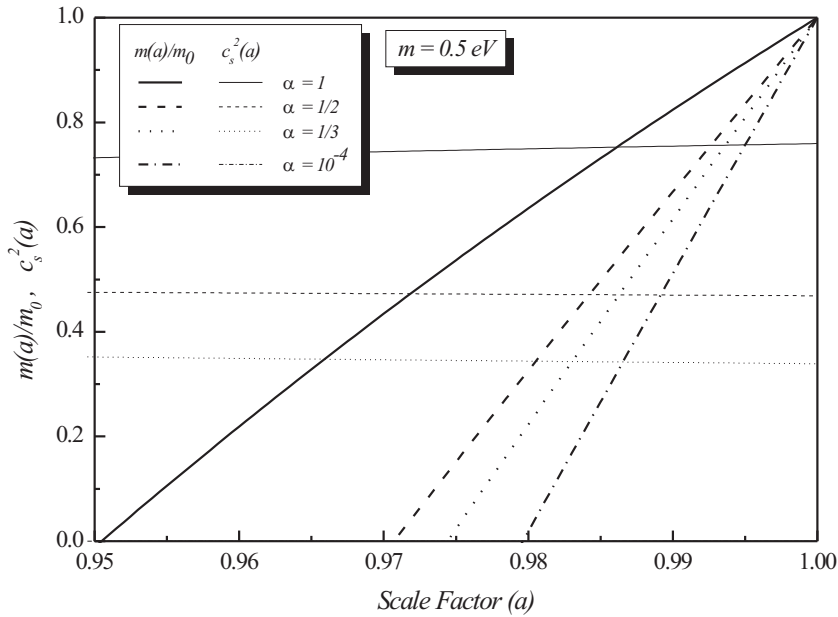}
\vspace{-0.5 cm}
\caption{Model independent perturbative modification on squared speed of sound $c^{2}_{s}$
as a function of the scale factor for the neutrino-GCG coupled fluid in comparison with the adiabatic
GCG fluid for $A_{s} = 0.7$ with $\alpha = 1, 1/2,\,1/3,\, 10^{-4}$, for a present-day value of neutrino mass, $m_{0} = 0.5\, eV$.}
\label{fGCG-10}
\end{figure}

Fig.~\ref{fGCG-10} illustrates the results for an increasing neutrino mass with the scale factor
for a set of phenomenologically consistent parameters in the context of the GCG model.
Interestingly, for $m_{0} = 0.5\, eV$, a fairly typical value, one can see that stable
MaVaN perturbations correspond to a well defined effective squared speed of sound,
\begin{equation}
c_{s}^{2} \simeq {d p_{\phi} \over d(\rho_{\phi}+ \rho_{\nu})} > 0.
\label{gcg33B}
\end{equation}
The greater the $m_{0}$ values, the more important are the corrections to the
squared speed of sound, up to the limit where the perturbative approach breaks down.
However, one finds that as far as the perturbative approach is concerned, our model does not run into
stability problems in the NR neutrino regime.
In opposition, in the SC treatment, where neutrinos are just coupled to dark energy, cosmic expansion
in combination with gravitational drag due to cold dark matter have a major impact on the stability of MaVaN models.
Usually, for a general fluid for which the equation of state is known, the dominant behaviour on $c_{s}^{2}$
arises from the dark sector component and not by the neutrino component.
For the models where the SC (cf. Eq.~(\ref{gcg07})) implies in a cosmological constant-type equation of
state, $ p_{\phi} = - \rho_{\phi}$, one inevitably obtains $c_{s}^{2} = -1$.

Thus, the perturbative approach is in agreement with the assumption that the coupling between neutrinos and
dark energy (and/or dark matter) is weak.
It is found that the stability condition related to the squared speed of sound of the coupled fluid is
predominantly governed by dark energy equation of state.
This is rather similar to the dynamics of the weakly coupled {\em cosmon} fields \cite{Wet94}.
Actually, such a troublesome behaviour should have already been observed as the SC implies that $c_{s}^{2} = -1$
from the very start, and the role of recovering the stability is relegated to the neutrino contribution \cite{Tak06}.
The loosening of the stationary constraint Eq. (\ref{gcg07}) emerges from the dynamical dependence on $\varphi$,
more concretely due to a kinetic energy component \cite{Ber08b}.
The knowledge of the background fluid equation of state for the dark sector (the GCG in our example), and the
criterion for the applicability of the perturbative approach, do allow to overcome the $c^2_{s}$ negative
problem, independently from the neutrino mass dependence set by the SC.


\section{Coupling to the Higgs field}

In order to study the coupling of dark energy to the Higgs field one considers first
the coupling of a very light scalar to the Higgs field. One examines a
model where the fields are already in their minimum energy configuration (contrary to the
usual quintessence models, where the fields are displaced from their minimum). As an
example one adopts the minimal spontaneously broken hidden sector model of Schabinger
and Wells \cite{wells} involving the usual Higgs doublet $H$ and a further singlet complex
scalar field $\Phi$ with a potential
\begin{equation}
V(\Phi,H) = -m_H^2 |H|^2 - m_\Phi^2 |\Phi|^2  + \lambda |H|^4 +
\rho |\Phi|^4  + \eta |H|^2 |\Phi|^2.
\end{equation}
The fields develop non-vanishing vacuum expectation values, $\langle |H|^2  \rangle =
v^2/2$ and $\langle |\Phi|^2  \rangle = \xi^2/2 $, and in unitary gauge, the physical
fields $h = H - v/\sqrt{2}$ (the Higgs boson field) and $\phi = \Phi - \xi/\sqrt{2}$ mix through a non-diagonal mass matrix due
to a non-vanishing $\eta$. The mixing angle $\omega$ between the two scalars is given by
\begin{equation}
\tan \omega = { \eta v \xi \over (\rho \xi^2 - \lambda v^2) \pm \sqrt{ ( \rho \xi^2 -
\lambda v^2)^2 + \eta^2 v^2 \xi^2} }~.
\end{equation}

In order to derive bounds on the mixing $\omega$ one considers fifth force constraints \cite{DZ}.
The mixing of the SM Higgs field with a light scalar induces a
long range Yukawa-like interaction due to the exchange of an ultra-light singlet field with
strength $g_e \sin \omega$ for electrons and $g_N \sin \omega$ for nucleons, where $g_e  =
2.9 \times 10^{-6}$ is the electron Yukawa coupling and $g_N = 2.1 \times 10^{-3}$ is the
nucleon Yukawa coupling \cite{cheng}.

Thus, a non-relativistic test body of mass $M$ placed in the
gravitational field of the Earth (with mass $M_E$) at a distance $r$ from its center
undergoes an acceleration given by:
\begin{equation}
a = a_{gr} + a_{\phi}~,
\end{equation}
where
\begin{equation}
a_{gr} = {M_E \over M_{Pl}^2 r^2}~,
\end{equation}
is the usual Newtonian acceleration, $M_{Pl}$ is the Planck mass and
\begin{equation}
a_{\phi} = {\omega^2 \over M r^2}[g_N^2  N_N^E N_N^t + g_N g_e (N_N^E N_e^t + N_e^E
N_N^t) + g_e^2 N_e^E N_e^t]
\end{equation}
is the extra acceleration due to a new force arising from $\phi$ exchange, as long as $m_\Phi r \ll 1$;
$ N_{(N,e)}^{(E,t)}$ is the number of nucleons or electrons in the Earth and in the
test body.

With these definitions one finds that the difference in acceleration between two test
bodies of distinct compositions is given by
\begin{equation}
\varepsilon = 2 {|a_1 - a_2| \over |a_1 + a_2|}= {M_{Pl} \over \bar{m}} \omega^2 g_N g_e \Delta f_p
\end{equation}
where $\bar{m}$ is the average nucleon mass and $\Delta f_p$ is the difference in isotopic composition
of the two test bodies:
\begin{equation}
\Delta f_p = {N_p^{(1)} \over N_p^{(1)} + N_n^{(1)}} - {N_p^{(2)} \over N_p^{(2)} +
N_n^{(2)}}
\end{equation}
with $N_{(p,n)}$ being the number of protons or neutrons in the test body.

For typical isotopic differences of the order of $\Delta f_p = {\cal O} (10^{-1})$ and
using the experimental result $\varepsilon < {\cal O} (10^{-13})$ \cite{baessler1,baessler2} one gets
an estimate for upper bound on the mixing angle
\begin{equation}
\omega < {\cal O} (10^{-20})~.
\end{equation}
Hence the mixing of an ultra-light
scalar singlet is severely constrained by fifth force-type experiments.
However, one should point out that this is
expected in the discussed model, since
in the limit $\xi \ll v$ the mixing angle is given by
\begin{equation}
|\omega| \simeq {\eta \over 2 \lambda} {\xi \over v} \simeq \eta {m_{\phi} \over m_H}
\end{equation}
where one assumes quartic couplings of $\cal{O}$(1). In order for this limit to
apply, the Compton length of the light scalar should be of the order of the Earth radius,
implying in  a mass of the order of $10^{-14}$ eV, to be compared with a Higgs mass of the
order of $100$ GeV. This leads to
$ \omega \simeq  \eta \times 10^{-25} $.
 Therefore, one concludes that there are no limits on the mixing constant
$\eta$, which implies that it is possible that the Higgs may have a large invisible width
into very light particles.

It is also possible that these light scalars are dark matter candidates. Since in
this case there is no symmetry preventing $\phi$ decay, one  must check whether in fact they
can survive till present. The light scalar decays through its mixing with the Higgs boson.
If $m_\phi < 2 m_e$, the dominant decay rate is into photons dominated by a top quark
triangle
($g_{\phi \bar{t} t} = \omega g_{h \bar{t} t}$), which can be readily estimated:
\begin{equation}
\Gamma(\phi \rightarrow \gamma \gamma) \simeq \omega^2 \left({\alpha \over \pi}\right)^2 {G_F
m_\phi^3 \over 36 \sqrt{2} \pi}
\end{equation}
which implies that
\begin{equation}
\Gamma(\phi \rightarrow \gamma \gamma)^{-1} \simeq {10^{-3} \over \omega^2 (m_\phi/\mbox{\footnotesize{1
eV}})^3}  t_U
\end{equation}
where $t_U \simeq 10^{17} s$ is the age of the universe. Hence, in order for the light scalar to survive till
present one must require
\begin{equation}
 \omega^2 \left({m_\phi \over 1 ~eV}\right)^3 < 10^{-3}
\end{equation}
which is easily satisfied for the considered case.

Hence one should consider the bounds discussed in Ref. \cite{bento2}, whose
reasoning shows that the scalar $\phi$ cannot be a dark matter candidate.
In order to do that one considers the cosmological evolution of the light scalar field. If
$\eta$ is sufficiently small, the singlet field decouples early in the thermal history of the
universe and is diluted by subsequent entropy production. In Ref.~\cite{bento1}, one has
considered out-of-equilibrium singlet production via inflaton decay in the context of
$N=1$ supergravity inflationary models (see. e.g. \cite{bento2} and references therein).

On the other hand, for certain values of the coupling $\eta$, it is possible that $\Phi$
particles are in thermal equilibrium with ordinary matter. In order to determine whether
this is the case, one makes the usual comparison between the thermalization rate
$\Gamma_{th}$ and the expansion rate of the universe, $H$.
In Ref.~\cite{bento1} it was found that if $\eta > 10^{-10}$ the singlets can be brought
into thermal equilibrium right after the electroweak phase transition, being therefore
as abundant as photons.

Since one is interested in a stable ultra-light singlet field, it is
a major concern avoiding its overproduction if it decouples while relativistic. In fact,
in this case, there is  an analogue of Lee-Weinberg limit for neutrinos (see e.g.
\cite{Kolb}):

\begin{equation}
\Omega_\phi h^2 \simeq 0.08 \left({m_\phi \over 1 ~ \ev}\right) \quad,
\end{equation}
which shows that as $m_{\phi} \simeq 10^{-14}$ eV, the thermally produced ultra-light
scalar $\phi$ particles cannot be a dark matter candidate, as for it is required that
$\Omega_\phi h^2 \simeq 0.1$. Of course, this model requires a considerable
fine-tuning to generate the different scalar mass scales and it should be embedded in a
more encompassing model, such as the Minimal Supersymmetric Standard Model with the
addition of one singlet chiral superfield \cite{BCFW}.

In what follows one considers the case where the Higgs is coupled to a quintessence field
and pay special attention to the
fact that it is not settled in its minimum. It is in this respect that a realistic dark
energy candidate differs from the generic dark matter candidate modelled by a
scalar $\Phi$ considered above .

The most natural potential coupling the Higgs field to the quintessence field
is of the ``hybrid inflation" type \cite{Linde}:
\begin{equation}
V(\Phi,H) =  U(\Phi) + \lambda \left( |H|^2 - {v^2 \over 2} \right)^2 +
\lambda_1 \Phi^2 |H|^2
\label{hybrid}
\end{equation}
where $U(\Phi)$ is the usual quintessential-type potential which can, for instance, be of the
form of an exponential or an inverse power law in the singlet field \cite{review1,review2}. The
second term is the SM Higgs potential and the last term gives rise to
Higgs-quintessence coupling, where $\lambda_1$ is a dimensionless coupling constant.

Working in the unitary gauge and denoting $\tilde{h}$ as the Higgs field, one can expand
the fields around their classical, homogeneous vacuum configurations:
\begin{equation}
\Phi(\vec{x},t) = \phi_c(t) + \varphi(\vec{x},t); \;\; \tilde{h}(\vec{x},t) = {1 \over \sqrt{2}} (\tilde{h}_c(t) + h(\vec{x},t))
\end{equation}
with $\langle \Phi(\vec{x},t) \rangle = \phi_c(t)$, $\langle \tilde{h}(\vec{x},t) \rangle = \tilde{h}_c(t)/\sqrt{2}$ and
the physical fields have zero vacuum expectation values,
$\langle \varphi(\vec{x},t) \rangle = \langle h(\vec{x},t)) \rangle = 0$.

The classical field configurations obey the equations of motion:
\begin{eqnarray}
\ddot{\phi}_{c} + 3 H \dot{\phi}_{c} + \left. {d U \over d \Phi} \right|_{\Phi = \phi_c} +  \lambda_1 \phi_c \tilde{h}_c^2 &=& 0 \nonumber \\
\ddot{\tilde{h}}_{c} + 3 H \dot{\tilde{h}}_{c} +  \lambda ( \tilde{h}_c^2 - v^2) \tilde{h}_c   +   2 \lambda_1 \phi_c^2 \tilde{h}_c &=& 0.
\end{eqnarray}
Since this is a model of dynamical dark energy, the classical quintessence field has not relaxed to the minimum
of its potential which causes the Higgs field to be displaced from the usual vacuum expectation value
($\tilde{h}_c \neq v$, where $v = 246$ GeV).
This results in a time-varying weak scale or $G_F$, for which there are very strong bounds \cite{varyG1,varyG2} (see also the bound Eq. (\ref{boundweak})).

In order to guarantee that $\tilde{h}_c = v$, one considers the following {\it ad hoc} modifications to the last term of the
potential Eq. (\ref{hybrid}), representing the interaction between the Higgs and quintessence fields:
\begin{eqnarray}
V(\Phi,H)_{int} = \lambda_1 \Phi^2 (|H| - v/\sqrt{2})^2 \;\; &(A)& \nonumber \\
V(\Phi,H)_{int} = \lambda_1 (\Phi - \phi_c)^2 |H|^2 \;\;  &(B)& \nonumber \\
 V(\Phi,H)_{int} = \lambda_1 (\Phi-\phi_c)^2 (|H| - v/\sqrt{2})^2 \;\; &(C)&
\label{potentials}
\end{eqnarray}

For case (A) the equations of motion become:
\begin{eqnarray}
\ddot{\phi}_{c} + 3 H \dot{\phi}_{c} + \left. {d U \over d \Phi} \right|_{\Phi = \phi_c} +  \lambda_1 \phi_c (\tilde{h}_c -v)^2 &=& 0 \nonumber \\
\ddot{\tilde{h}}_{c} + 3 H \dot{\tilde{h}}_{c} +  \lambda ( \tilde{h}_c^2 - v^2) \tilde{h}_c   +   \sqrt{2} \lambda_1 \phi_c^2 (\tilde{h}_c - v) &=& 0 \nonumber \\
\end{eqnarray}
and hence $\tilde{h}_c = v$ is a solution in which case the classical quintessence field obeys the usual equation:
\begin{equation}
\ddot{\phi}_{c} + 3 H \dot{\phi}_{c} + \left. {d U \over d \Phi} \right|_{\Phi = \phi_c}  = 0.
\label{ev}
\end{equation}
It is interesting to notice that this vacuum state for the $\Phi$ field
does not arise from a mechanism of spontaneous
symmetry breaking, but is instead a consequence of initial conditions for the slow roll
evolution of the quintessence field. For most of the models $\phi_{c}$ evolves essentially
as a logarithm of the cosmic time, so that its typical rate of change is the Hubble time:
\begin{equation}
{\dot{\phi}_{c} \over \phi_{c}} \sim H ~.
\end{equation}
Thus, $\phi_{c}$ evolves in a quasi-stationary regime set by the Hubble scale.

The Lagrangian density for the physical fields is given by
\begin{equation}
{\cal L} = {1 \over 2} \left[ ( \dot{\phi}_{c} + \dot{\varphi})^2 - (\vec{\nabla} \varphi)^2 \right] + {1 \over 2} (\partial_\mu h)^2 -
V(\phi_c,v,\varphi,h)
\label{L1}
\end{equation}
where
\begin{eqnarray}
V(\phi_c,v,\varphi,h) &=& U(\phi_c + \varphi) + \lambda \left( {1 \over 2}(h+v)^2 - v^2/2 \right)^2 + \nonumber \\
&& \frac{1}{2}\lambda_1 (\phi_c + \varphi)^2 h^2.
\end{eqnarray}
Writing
\begin{equation}
U(\phi_c + \varphi)  = U(\phi_c) + U^\prime(\phi_c) \varphi + \frac{1}{2} U^{\prime\prime}(\phi_c) \varphi^2 + \frac{1}{6} U^{\prime\prime\prime}(\phi_c) \varphi^3 + \cdots
\label{U}
\end{equation}
and using the classical equation of motion to cancel the tadpole in Eq. (\ref{U}) with the $\dot{\phi}_{c} \dot{\varphi}$
term in Eq. (\ref{L1}) one finally obtains:
\begin{eqnarray}
{\cal L} &=& \left[ \frac{1}{2}  \dot{\phi}_{c}^2 -  U(\phi_c) \right] + \frac{1}{2} (\partial_\mu \varphi)^2 +
\frac{1}{2} (\partial_\mu h)^2 - \frac{1}{2} m_\varphi^2 \varphi^2 - \nonumber \\
&& \lambda (h^2/2 +  v h)^2  - \frac{1}{2} \lambda_1 (\phi_c + \varphi)^2 h^2 - \frac{1}{6} g \varphi^3 + {\cal O}(\varphi^4) \nonumber \\
\label{L2}
\end{eqnarray}
where $m_\varphi^2 = U^{\prime\prime}(\phi_c)$ and $g = U^{\prime\prime\prime}(\phi_c)$.

The term in brackets in Eq. (\ref{L2}) reflects the classical zero mode contribution of the scalar field to
the energy density of the universe which causes its accelerated expansion.
The other terms express the behaviour of the quantum excitations (particles) in the universe. In particular,
the Higgs-quintessence interaction results in a new, time-dependent contribution to the Higgs boson mass given by:
\begin{equation}
m_h^2 =  2 \lambda v^2 + \lambda_1 \phi_c^2~.
\end{equation}
Since in usual quintessence models $\phi_c = {\cal O}(M_{Pl})$ one expects a very large deviation from
the Standard Model value of the Higgs mass unless the coupling constant $\lambda_1$
is very tiny, $\lambda_1 = {\cal O} (v^2/M_{Pl}^2)$ practically closing the Higgs portal.
Notice, however, that unitarity in longitudinal gauge boson scattering is not violated since
the relevant couplings are still small even if the Higgs boson mass is large.

In case (B) the only difference arises in the interaction term, which becomes
\begin{equation}
{\cal L}_{int} = - \frac{1}{2} \lambda_1\varphi^2 (v+h)^2
\end{equation}
and there is no new contribution to the Higgs mass. Interestingly, in this case the coupling
generates a new contribution to the mass of the quintessence particle $\varphi$:
\begin{equation}
m_\varphi^2 =  U^{\prime\prime}(\phi_c) + \lambda_1 v^2
\end{equation}
that, contrary to usual quintessence models, does not have to be small, since it does not
affect the evolution of $\phi_{c}(t)$ given by Eq. (\ref{ev}).
Furthermore, since the original quintessence field is not
coupled to ordinary matter and there is no mixing to the Higgs boson, there are no
bounds arising from fifth force constraints. Notice that in strict terms, this choice renders an ill defined field theory as
the $\varphi$ mass will vary on a cosmological scale. However, for models where the acceleration of the expansion is transient,
it is possible that $U^{\prime\prime}(\phi_c)$ vanishes around present time (see e.g. Refs. \cite{Skordis1,Skordis2,Bento02}).
Under this condition,
this case also introduces a trilinear coupling $\varphi^2 h$ which, in addition to generating an invisible Higgs boson decay,
can bring the quintessence particle into thermal equilibrium in the early universe.
The analysis of the contribution arising from this quintessence particle to the dark matter component
follows that of Ref. \cite{mcdonald,bento2,burgess}. An interesting situation occurs for the natural
case of $\lambda_1 = {\cal O} (0.04)$, which implies $m_\varphi \simeq 50$ GeV.
In this case, the $\varphi$ particle can be brought into thermal equilibrium through the Higgs
coupling in the early universe and, for a Higgs mass $m_H \simeq 140$ GeV it  decouples while non-relativistic,
generating the dark matter density in the observed range.

Therefore, in this unified picture, dark energy is the zero-mode classical field $\phi_{c}(t)$
rolling down the usual quintessence potential and the dark matter candidate is the quantum
excitation (particle) $\varphi$, which is produced in the universe due to its coupling to
the Higgs boson.

Finally, in case (C) there are no new contributions to either to the Higgs or to the quintessence particle.
Only a quartic interaction $\varphi^2 h^2$ is generated and the quintessence particle turns out not to be a good dark matter
candidate.


\section{Conclusions}

In this contribution, the implications of coupling dark energy to SM fields, namely
gauge fields, neutrinos and the Higgs field, are examined.
One has first considered a possible  coupling of dark energy
to gauge fields and the impact it might have on the nuclear decay Rutherford-Soddy law. From the
available bounds on the variation of the strong and weak gauge couplings it has been
estimated, the deviations from the Rutherford-Soddy law. It has been considered a linear model for the variation of the couplings.
One finds that the stringent bound on
the variation of the weak coupling arising from $^{187}Re$ implies in
significant constraints on any
variation of the Rutherford-Soddy law for weak interactions,
actually at the level $3 \times 10^{-7}$ for
the quintessence field with an exponential potential and at $8 \times 10^{-7}$ level for the GCG model.
For strong interactions, the bounds are much less stringent and are
about $4 \times 10^{-2}$ for the quintessence
model with exponential potential and of order $10^{-1}$ for the GCG model.
Stronger
bounds on the deviation of the nuclear decay law for strong interactions could be obtained if
variations of the strong coupling were tighter. Furthermore, it is
interesting that somewhat different deviations are found for distinct
dark energy models. This may be relevant to distinguish them given that most
often they are degenerate with respect to the observable cosmological parameters.

In what concerns the coupling of neutrinos to dark energy, one finds that models with increasing
neutrino mass with the scale factor is phenomenologically consistent in the context of the GCG model.
For $m_{0} = 0.5\, eV$, a fairly typical value, one can see that stable MaVaN perturbations correspond to
a well defined effective squared speed of sound.
The greater  the $m_{0}$ values are, the more important are the corrections to the squared speed of sound,
up to the limit where the perturbative approach breaks down \cite{Ber08b}.

As for the coupling of dark energy to the Higgs field, the implications of a scenario where the
Higgs boson is coupled to a SM singlet field responsible for the accelerated expansion of
the universe were examined. The most natural ``hybrid-like" potential is disfavoured since it leads
to a time-variation of the weak scale.
A modification of the potential can however, give origin to
the quite interesting possibility where the classical
zero-mode component of the singlet field corresponds to the dark energy particle
while its excitation
plays the role of dark matter. In order to make
this excitation consistent with
the cosmological density requirement $\Omega_\varphi h^2 \simeq 0.1$ one gets a coupling
to the Higgs field $\lambda_1 = {\cal O} (10^{-2})$, which implies   $m_\varphi =  {\cal O} (10)$ GeV.
Quite interestingly, this scenario can be  be scrutinized in the forthcoming generation of
accelerators through the invisible decay of the Higgs boson \cite{rosen1}.

\ack

O.B would like to thank R. Rosenfeld for collaboration on the work reported on Section 4. He would also like to thank
T. Elze for setting up so successfully DICE 2008 and for creating such a nice atmosphere for discussion and fruitful exchange of ideas.

\noindent


\end{document}